\documentclass [12pt]{article}
\usepackage{a4}
\usepackage{epsf}
\usepackage{epsfig}
\def\invpb{\ensuremath{{{\rm pb}^{-1}}}}       
       
\def\G{\ensuremath{{\mathrm{GeV}}}}
\def\Gc{\ensuremath{{\mathrm{GeV}/c}}}
\def\Gcs{\ensuremath{{\mathrm{GeV}/c^2}}}

\def\h{\ensuremath{{\mathrm h}}}
\def\A{\ensuremath{{\mathrm A}}}
\def\Z{\ensuremath{{\mathrm Z}}}
\def\W{\ensuremath{{\mathrm W}}}
\def\e{\ensuremath{{\mathrm e}}}
\def\l{\ensuremath{{\ell}}}

\def\WW{\ensuremath{{\mathrm W}^{+}{\mathrm W}^{-}}}
\def\ZZ{\ensuremath{{\mathrm Z}{\mathrm Z}}}
\def\ee{\ensuremath{{\mathrm e}^{+}{\mathrm e}^{-}}}
\def\epem{\ee}
\def\ffbar{\ensuremath{{\mathrm f}\bar{\mathrm f}}}
\def\bb{\ensuremath{{\mathrm b}\bar{\mathrm b}}}
\def\bbbar{\bb}
\def\qq{\ensuremath{{\mathrm q}\bar{\mathrm q}}}
\def\qqbar{\qq}

\def\nunu{\mbox{\ensuremath{\nu\bar{\nu}}}}

\def\tptm{\mbox{\ensuremath{{\tau}^{+}{\tau}^{-}}}}
\def\mpmm{\mbox{\ensuremath{{\mu}^{+}{\mu}^{-}}}}
\def\lplm{\mbox{\ensuremath{{\l}^{+}{\l}^{-}}}}

\def\mh{\mbox{\ensuremath{m_{\h}}}}

\def\mA{\mbox{\ensuremath{m_{\A}}}}

\newcommand{\MSSM}{MSSM}
\def\to{\mbox{\ensuremath{ \rightarrow \ }}}

\def\tanb{\ensuremath{\tan\beta}}
\def\sba{\ensuremath{\sin^2(\beta-\alpha)}}
\def\sinba{\sba}
\def\cba{\ensuremath{\cos^2(\beta-\alpha)}}
\def\pt{\ensuremath{p_{\mathrm{T}}}}

\newcommand{\ALEPH}{{ALEPH}}

\newcommand{\LEP}{{LEP}}

\newcommand{\CERN}{{CERN}}
\def\capstyl{\small}
\def\PLB#1#2#3{{ Phys. Lett. }{\bf B#1 }(#2) #3}
\def\NPB#1#2#3{{ Nucl. Phys. }{\bf B#1 }(#2) #3}

\def\EPJ#1#2#3{{ E. Phys. J. }{\bf C#1 }(#2) #3}
\def\NIM#1#2#3{{ Nucl. Instrum. and Methods}
{\bf A#1 }(#2) #3}
\def\CPC#1#2#3{{ Comput. Phys. Commun. }{\bf #1 }(#2) #3}
\def\etal{{\it et al.}}
\def\PYTH{{\sc pythia}}
\def\KORALZ{{\sc koralz}}

\def\HZHA{{\sc hzha}}
\def\GRACE{{\sc grace4f}}

\begin{document}
\begin{titlepage}
\begin{flushright}
CERN-EP/2000-131\\
17 October 2000\\
\end{flushright}
\setlength{\topmargin}{0.5cm}
\setlength{\oddsidemargin}{-0.2cm}
\vspace{1cm}
\begin{center}
\begin{LARGE}
{Searches for neutral Higgs bosons in e$^+$e$^-$ collisions\\ 
at centre-of-mass energies from 192 to 202 GeV\\}
\end{LARGE}
\end{center}
\vskip 1cm
\begin{center}
\begin{large}
{The {\sc ALEPH} Collaboration}$^{*)}$
\end{large}
\end{center}
\vspace{1cm}
\thispagestyle{empty}
\begin{picture}(160,1)
\put(-3, 110){\rm\Large EUROPEAN ORGANIZATION FOR NUCLEAR RESEARCH (CERN)}
\end{picture}

\begin{abstract}
\noindent 
Searches for neutral Higgs bosons are performed with the $237\,\invpb$
of data collected in 1999 by the ALEPH detector at LEP, for
centre-of-mass energies between 191.6 and $201.6\,\G$.  These searches
apply to Higgs bosons within the context of the Standard Model and its
minimal supersymmetric extension (\MSSM) as well as to invisibly
decaying Higgs bosons.  No evidence of a signal is seen.  A lower
limit on the mass of the Standard Model Higgs boson of 107.7\,\Gcs\ at
$95\%$~confidence level is set.  In the \MSSM, lower limits of 91.2
and 91.6\,\Gcs\ are derived for the masses of the neutral Higgs bosons
\h\ and A, respectively.  For a Higgs boson decaying invisibly and produced with the
Standard Model cross section, masses below 106.4\,\Gcs\ are excluded.
\end{abstract}
\vspace{0.2cm}
\begin{center}
({\sl To be submitted to Physics Letters})
\end{center}
\vspace{0.4cm}
\begin{center}
*) See next pages for the list of authors.
\end{center}

\end{titlepage}

\newpage
%
\pagestyle{empty}
\newpage
\small
%
%
\newlength{\saveparskip}
\newlength{\savetextheight}
\newlength{\savetopmargin}
\newlength{\savetextwidth}
\newlength{\saveoddsidemargin}
\newlength{\savetopsep}
\setlength{\saveparskip}{\parskip}
\setlength{\savetextheight}{\textheight}
\setlength{\savetopmargin}{\topmargin}
\setlength{\savetextwidth}{\textwidth}
\setlength{\saveoddsidemargin}{\oddsidemargin}
\setlength{\savetopsep}{\topsep}
%
%
\setlength{\parskip}{0.0cm}
\setlength{\textheight}{25.0cm}
\setlength{\topmargin}{-1.5cm}
\setlength{\textwidth}{16 cm}
\setlength{\oddsidemargin}{-0.0cm}
\setlength{\topsep}{1mm}
\pretolerance=10000
\centerline{\large\bf The ALEPH Collaboration}
\footnotesize
\vspace{0.5cm}
{\raggedbottom
\begin{sloppypar}
\samepage\noindent
R.~Barate,
D.~Decamp,
P.~Ghez,
C.~Goy,
S.~Jezequel,
J.-P.~Lees,
F.~Martin,
E.~Merle,
\mbox{M.-N.~Minard},
B.~Pietrzyk
\nopagebreak
\begin{center}
\parbox{15.5cm}{\sl\samepage
Laboratoire de Physique des Particules (LAPP), IN$^{2}$P$^{3}$-CNRS,
F-74019 Annecy-le-Vieux Cedex, France}
\end{center}\end{sloppypar}
\vspace{2mm}
\begin{sloppypar}
\noindent
S.~Bravo,
M.P.~Casado,
M.~Chmeissani,
J.M.~Crespo,
E.~Fernandez,
M.~Fernandez-Bosman,
Ll.~Garrido,$^{15}$
E.~Graug\'{e}s,
J.~Lopez,
M.~Martinez,
G.~Merino,
R.~Miquel,
Ll.M.~Mir,
A.~Pacheco,
D.~Paneque,
H.~Ruiz
\nopagebreak
\begin{center}
\parbox{15.5cm}{\sl\samepage
Institut de F\'{i}sica d'Altes Energies, Universitat Aut\`{o}noma
de Barcelona, E-08193 Bellaterra (Barcelona), Spain$^{7}$}
\end{center}\end{sloppypar}
\vspace{2mm}
\begin{sloppypar}
\noindent
A.~Colaleo,
D.~Creanza,
N.~De~Filippis,
M.~de~Palma,
G.~Iaselli,
G.~Maggi,
M.~Maggi,$^{1}$
S.~Nuzzo,
A.~Ranieri,
G.~Raso,$^{24}$
F.~Ruggieri,
G.~Selvaggi,
L.~Silvestris,
P.~Tempesta,
A.~Tricomi,$^{3}$
G.~Zito
\nopagebreak
\begin{center}
\parbox{15.5cm}{\sl\samepage
Dipartimento di Fisica, INFN Sezione di Bari, I-70126 Bari, Italy}
\end{center}\end{sloppypar}
\vspace{2mm}
\begin{sloppypar}
\noindent
X.~Huang,
J.~Lin,
Q. Ouyang,
T.~Wang,
Y.~Xie,
R.~Xu,
S.~Xue,
J.~Zhang,
L.~Zhang,
W.~Zhao
\nopagebreak
\begin{center}
\parbox{15.5cm}{\sl\samepage
Institute of High Energy Physics, Academia Sinica, Beijing, The People's
Republic of China$^{8}$}
\end{center}\end{sloppypar}
\vspace{2mm}
\begin{sloppypar}
\noindent
D.~Abbaneo,
P.~Azzurri,
G.~Boix,$^{6}$
O.~Buchm\"uller,
M.~Cattaneo,
F.~Cerutti,
B.~Clerbaux,
G.~Dissertori,
H.~Drevermann,
R.W.~Forty,
M.~Frank,
F.~Gianotti,
T.C.~Greening,
J.B.~Hansen,
J.~Harvey,
D.E.~Hutchcroft,
P.~Janot,
B.~Jost,
M.~Kado,
V.~Lemaitre,
P.~Maley,
P.~Mato,
A.~Minten,
A.~Moutoussi,
F.~Ranjard,
L.~Rolandi,
D.~Schlatter,
M.~Schmitt,$^{20}$
O.~Schneider,$^{2}$
P.~Spagnolo,
W.~Tejessy,
F.~Teubert,
E.~Tournefier,$^{26}$
A.~Valassi,
J.J.~Ward,
A.E.~Wright
\nopagebreak
\begin{center}
\parbox{15.5cm}{\sl\samepage
European Laboratory for Particle Physics (CERN), CH-1211 Geneva 23,
Switzerland}
\end{center}\end{sloppypar}
\vspace{2mm}
\begin{sloppypar}
\noindent
Z.~Ajaltouni,
F.~Badaud,
S.~Dessagne,
A.~Falvard,
D.~Fayolle,
P.~Gay,
P.~Henrard,
J.~Jousset,
B.~Michel,
S.~Monteil,
\mbox{J-C.~Montret},
D.~Pallin,
J.M.~Pascolo,
P.~Perret,
F.~Podlyski
\nopagebreak
\begin{center}
\parbox{15.5cm}{\sl\samepage
Laboratoire de Physique Corpusculaire, Universit\'e Blaise Pascal,
IN$^{2}$P$^{3}$-CNRS, Clermont-Ferrand, F-63177 Aubi\`{e}re, France}
\end{center}\end{sloppypar}
\vspace{2mm}
\begin{sloppypar}
\noindent
J.D.~Hansen,
J.R.~Hansen,
P.H.~Hansen,
B.S.~Nilsson,
A.~W\"a\"an\"anen
\nopagebreak
\begin{center}
\parbox{15.5cm}{\sl\samepage
Niels Bohr Institute, 2100 Copenhagen, DK-Denmark$^{9}$}
\end{center}\end{sloppypar}
\vspace{2mm}
\begin{sloppypar}
\noindent
G.~Daskalakis,
A.~Kyriakis,
C.~Markou,
E.~Simopoulou,
A.~Vayaki
\nopagebreak
\begin{center}
\parbox{15.5cm}{\sl\samepage
Nuclear Research Center Demokritos (NRCD), GR-15310 Attiki, Greece}
\end{center}\end{sloppypar}
\vspace{2mm}
\begin{sloppypar}
\noindent
A.~Blondel,$^{12}$
\mbox{J.-C.~Brient},
F.~Machefert,
A.~Roug\'{e},
M.~Swynghedauw,
R.~Tanaka
\linebreak
H.~Videau
\nopagebreak
\begin{center}
\parbox{15.5cm}{\sl\samepage
Laboratoire de Physique Nucl\'eaire et des Hautes Energies, Ecole
Polytechnique, IN$^{2}$P$^{3}$-CNRS, \mbox{F-91128} Palaiseau Cedex, France}
\end{center}\end{sloppypar}
\vspace{2mm}
\begin{sloppypar}
\noindent
E.~Focardi,
G.~Parrini,
K.~Zachariadou
\nopagebreak
\begin{center}
\parbox{15.5cm}{\sl\samepage
Dipartimento di Fisica, Universit\`a di Firenze, INFN Sezione di Firenze,
I-50125 Firenze, Italy}
\end{center}\end{sloppypar}
\vspace{2mm}
\begin{sloppypar}
\noindent
A.~Antonelli,
M.~Antonelli,
G.~Bencivenni,
G.~Bologna,$^{4}$
F.~Bossi,
P.~Campana,
G.~Capon,
V.~Chiarella,
P.~Laurelli,
G.~Mannocchi,$^{5}$
F.~Murtas,
G.P.~Murtas,
L.~Passalacqua,
M.~Pepe-Altarelli$^{25}$
\nopagebreak
\begin{center}
\parbox{15.5cm}{\sl\samepage
Laboratori Nazionali dell'INFN (LNF-INFN), I-00044 Frascati, Italy}
\end{center}\end{sloppypar}
\vspace{2mm}
\begin{sloppypar}
\noindent
M.~Chalmers,
A.W.~Halley,
J.~Kennedy,
J.G.~Lynch,
P.~Negus,
V.~O'Shea,
B.~Raeven,
D.~Smith,
P.~Teixeira-Dias,
A.S.~Thompson
\nopagebreak
\begin{center}
\parbox{15.5cm}{\sl\samepage
Department of Physics and Astronomy, University of Glasgow, Glasgow G12
8QQ,United Kingdom$^{10}$}
\end{center}\end{sloppypar}
\begin{sloppypar}
\noindent
R.~Cavanaugh,
S.~Dhamotharan,
C.~Geweniger,
P.~Hanke,
V.~Hepp,
E.E.~Kluge,
G.~Leibenguth,
A.~Putzer,
K.~Tittel,
S.~Werner,$^{19}$
M.~Wunsch$^{19}$
\nopagebreak
\begin{center}
\parbox{15.5cm}{\sl\samepage
Kirchhoff-Institut f\"ur Physik, Universit\"at Heidelberg, D-69120
Heidelberg, Germany$^{16}$}
\end{center}\end{sloppypar}
\vspace{2mm}
\begin{sloppypar}
\noindent
R.~Beuselinck,
D.M.~Binnie,
W.~Cameron,
G.~Davies,
P.J.~Dornan,
M.~Girone,$^{1}$
N.~Marinelli,
J.~Nowell,
H.~Przysiezniak,
J.K.~Sedgbeer,
J.C.~Thompson,$^{14}$
E.~Thomson,$^{23}$
R.~White
\nopagebreak
\begin{center}
\parbox{15.5cm}{\sl\samepage
Department of Physics, Imperial College, London SW7 2BZ,
United Kingdom$^{10}$}
\end{center}\end{sloppypar}
\vspace{2mm}
\begin{sloppypar}
\noindent
V.M.~Ghete,
P.~Girtler,
E.~Kneringer,
D.~Kuhn,
G.~Rudolph
\nopagebreak
\begin{center}
\parbox{15.5cm}{\sl\samepage
Institut f\"ur Experimentalphysik, Universit\"at Innsbruck, A-6020
Innsbruck, Austria$^{18}$}
\end{center}\end{sloppypar}
\vspace{2mm}
\begin{sloppypar}
\noindent
E.~Bouhova-Tracker,
C.K.~Bowdery,
P.G.~Buck,
D.P.~Clarke,
G.~Ellis,
A.J.~Finch,
F.~Foster,
G.~Hughes,
R.W.L.~Jones,$^{1}$
N.A.~Robertson,
M.~Smizanska
\nopagebreak
\begin{center}
\parbox{15.5cm}{\sl\samepage
Department of Physics, University of Lancaster, Lancaster LA1 4YB,
United Kingdom$^{10}$}
\end{center}\end{sloppypar}
\vspace{2mm}
\begin{sloppypar}
\noindent
I.~Giehl,
F.~H\"olldorfer,
K.~Jakobs,
K.~Kleinknecht,
M.~Kr\"ocker,
A.-S.~M\"uller,
H.-A.~N\"urnberger,
G.~Quast,$^{1}$
B.~Renk,
E.~Rohne,
H.-G.~Sander,
S.~Schmeling,
H.~Wachsmuth,
C.~Zeitnitz,
T.~Ziegler
\nopagebreak
\begin{center}
\parbox{15.5cm}{\sl\samepage
Institut f\"ur Physik, Universit\"at Mainz, D-55099 Mainz, Germany$^{16}$}
\end{center}\end{sloppypar}
\vspace{2mm}
\begin{sloppypar}
\noindent
A.~Bonissent,
J.~Carr,
P.~Coyle,
C.~Curtil,
A.~Ealet,
D.~Fouchez,
O.~Leroy,
T.~Kachelhoffer,
P.~Payre,
D.~Rousseau,
A.~Tilquin
\nopagebreak
\begin{center}
\parbox{15.5cm}{\sl\samepage
Centre de Physique des Particules de Marseille, Univ M\'editerran\'ee,
IN$^{2}$P$^{3}$-CNRS, F-13288 Marseille, France}
\end{center}\end{sloppypar}
\vspace{2mm}
\begin{sloppypar}
\noindent
M.~Aleppo,
S.~Gilardoni,
F.~Ragusa
\nopagebreak
\begin{center}
\parbox{15.5cm}{\sl\samepage
Dipartimento di Fisica, Universit\`a di Milano e INFN Sezione di
Milano, I-20133 Milano, Italy.}
\end{center}\end{sloppypar}
\vspace{2mm}
\begin{sloppypar}
\noindent
A.~David,
H.~Dietl,
G.~Ganis,$^{27}$
K.~H\"uttmann,
G.~L\"utjens,
C.~Mannert,
W.~M\"anner,
\mbox{H.-G.~Moser},
S.~Schael,
R.~Settles,$^{1}$
H.~Stenzel,
W.~Wiedenmann,
G.~Wolf
\nopagebreak
\begin{center}
\parbox{15.5cm}{\sl\samepage
Max-Planck-Institut f\"ur Physik, Werner-Heisenberg-Institut,
D-80805 M\"unchen, Germany\footnotemark[16]}
\end{center}\end{sloppypar}
\vspace{2mm}
\begin{sloppypar}
\noindent
J.~Boucrot,$^{1}$
O.~Callot,
M.~Davier,
L.~Duflot,
\mbox{J.-F.~Grivaz},
Ph.~Heusse,
A.~Jacholkowska,$^{1}$
L.~Serin,
\mbox{J.-J.~Veillet},
I.~Videau,
J.-B.~de~Vivie~de~R\'egie,
C.~Yuan,
D.~Zerwas
\nopagebreak
\begin{center}
\parbox{15.5cm}{\sl\samepage
Laboratoire de l'Acc\'el\'erateur Lin\'eaire, Universit\'e de Paris-Sud,
IN$^{2}$P$^{3}$-CNRS, F-91898 Orsay Cedex, France}
\end{center}\end{sloppypar}
\vspace{2mm}
\begin{sloppypar}
\noindent
G.~Bagliesi,
T.~Boccali,
G.~Calderini,
V.~Ciulli,
L.~Fo\`a,
A.~Giammanco,
A.~Giassi,
F.~Ligabue,
A.~Messineo,
F.~Palla,$^{1}$
G.~Sanguinetti,
A.~Sciab\`a,
G.~Sguazzoni,
R.~Tenchini,$^{1}$
A.~Venturi,
P.G.~Verdini
\samepage
\begin{center}
\parbox{15.5cm}{\sl\samepage
Dipartimento di Fisica dell'Universit\`a, INFN Sezione di Pisa,
e Scuola Normale Superiore, I-56010 Pisa, Italy}
\end{center}\end{sloppypar}
\vspace{2mm}
\begin{sloppypar}
\noindent
G.A.~Blair,
J.~Coles,
G.~Cowan,
M.G.~Green,
L.T.~Jones,
T.~Medcalf,
J.A.~Strong,
\mbox{J.H.~von~Wimmersperg-Toeller} 
\nopagebreak
\begin{center}
\parbox{15.5cm}{\sl\samepage
Department of Physics, Royal Holloway \& Bedford New College,
University of London, Surrey TW20 OEX, United Kingdom$^{10}$}
\end{center}\end{sloppypar}
\vspace{2mm}
\begin{sloppypar}
\noindent
R.W.~Clifft,
T.R.~Edgecock,
P.R.~Norton,
I.R.~Tomalin
\nopagebreak
\begin{center}
\parbox{15.5cm}{\sl\samepage
Particle Physics Dept., Rutherford Appleton Laboratory,
Chilton, Didcot, Oxon OX11 OQX, United Kingdom$^{10}$}
\end{center}\end{sloppypar}
\vspace{2mm}
\begin{sloppypar}
\noindent
\mbox{B.~Bloch-Devaux},$^{1}$
D.~Boumediene,
P.~Colas,
B.~Fabbro,
E.~Lan\c{c}on,
\mbox{M.-C.~Lemaire},
E.~Locci,
P.~Perez,
J.~Rander,
\mbox{J.-F.~Renardy},
A.~Rosowsky,
P.~Seager,$^{13}$
A.~Trabelsi,$^{21}$
B.~Tuchming,
B.~Vallage
\nopagebreak
\begin{center}
\parbox{15.5cm}{\sl\samepage
CEA, DAPNIA/Service de Physique des Particules,
CE-Saclay, F-91191 Gif-sur-Yvette Cedex, France$^{17}$}
\end{center}\end{sloppypar}
\vspace{2mm}
\begin{sloppypar}
\noindent
N.~Konstantinidis,
C.~Loomis,
A.M.~Litke,
G.~Taylor
\nopagebreak
\begin{center}
\parbox{15.5cm}{\sl\samepage
Institute for Particle Physics, University of California at
Santa Cruz, Santa Cruz, CA 95064, USA$^{22}$}
\end{center}\end{sloppypar}
\vspace{2mm}
\begin{sloppypar}
\noindent
C.N.~Booth,
S.~Cartwright,
F.~Combley,
P.N.~Hodgson,
M.~Lehto,
L.F.~Thompson
\nopagebreak
\begin{center}
\parbox{15.5cm}{\sl\samepage
Department of Physics, University of Sheffield, Sheffield S3 7RH,
United Kingdom$^{10}$}
\end{center}\end{sloppypar}
\vspace{2mm}
\begin{sloppypar}
\noindent
K.~Affholderbach,
A.~B\"ohrer,
S.~Brandt,
C.~Grupen,
J.~Hess,
A.~Misiejuk,
G.~Prange,
U.~Sieler
\nopagebreak
\begin{center}
\parbox{15.5cm}{\sl\samepage
Fachbereich Physik, Universit\"at Siegen, D-57068 Siegen, Germany$^{16}$}
\end{center}\end{sloppypar}
\vspace{2mm}
\begin{sloppypar}
\noindent
C.~Borean,
G.~Giannini,
B.~Gobbo
\nopagebreak
\begin{center}
\parbox{15.5cm}{\sl\samepage
Dipartimento di Fisica, Universit\`a di Trieste e INFN Sezione di Trieste,
I-34127 Trieste, Italy}
\end{center}\end{sloppypar}
\vspace{2mm}
\begin{sloppypar}
\noindent
H.~He,
J.~Putz,
J.~Rothberg,
S.~Wasserbaech
\nopagebreak
\begin{center}
\parbox{15.5cm}{\sl\samepage
Experimental Elementary Particle Physics, University of Washington, Seattle,
WA 98195 U.S.A.}
\end{center}\end{sloppypar}
\vspace{2mm}
\begin{sloppypar}
\noindent
S.R.~Armstrong,
K.~Cranmer,
P.~Elmer,
D.P.S.~Ferguson,
Y.~Gao,
S.~Gonz\'{a}lez,
O.J.~Hayes,
H.~Hu,
S.~Jin,
J.~Kile,
P.A.~McNamara III,
J.~Nielsen,
W.~Orejudos,
Y.B.~Pan,
Y.~Saadi,
I.J.~Scott,
J.~Walsh,
J.~Wu,
Sau~Lan~Wu,
X.~Wu,
G.~Zobernig
\nopagebreak
\begin{center}
\parbox{15.5cm}{\sl\samepage
Department of Physics, University of Wisconsin, Madison, WI 53706,
USA$^{11}$}
\end{center}\end{sloppypar}
}
\footnotetext[1]{Also at CERN, 1211 Geneva 23, Switzerland.}
\footnotetext[2]{Now at Universit\'e de Lausanne, 1015 Lausanne, Switzerland.}
\footnotetext[3]{Also at Dipartimento di Fisica di Catania and INFN Sezione di
 Catania, 95129 Catania, Italy.}
\footnotetext[4]{Deceased.}
\footnotetext[5]{Also Istituto di Cosmo-Geofisica del C.N.R., Torino,
Italy.}
\footnotetext[6]{Supported by the Commission of the European Communities,
contract ERBFMBICT982894.}
\footnotetext[7]{Supported by CICYT, Spain.}
\footnotetext[8]{Supported by the National Science Foundation of China.}
\footnotetext[9]{Supported by the Danish Natural Science Research Council.}
\footnotetext[10]{Supported by the UK Particle Physics and Astronomy Research
Council.}
\footnotetext[11]{Supported by the US Department of Energy, grant
DE-FG0295-ER40896.}
\footnotetext[12]{Now at Departement de Physique Corpusculaire, Universit\'e de
Gen\`eve, 1211 Gen\`eve 4, Switzerland.}
\footnotetext[13]{Supported by the Commission of the European Communities,
contract ERBFMBICT982874.}
\footnotetext[14]{Also at Rutherford Appleton Laboratory, Chilton, Didcot, UK.}
\footnotetext[15]{Permanent address: Universitat de Barcelona, 08208 Barcelona,
Spain.}
\footnotetext[16]{Supported by the Bundesministerium f\"ur Bildung,
Wissenschaft, Forschung und Technologie, Germany.}
\footnotetext[17]{Supported by the Direction des Sciences de la
Mati\`ere, C.E.A.}
\footnotetext[18]{Supported by the Austrian Ministry for Science and Transport.}
\footnotetext[19]{Now at SAP AG, 69185 Walldorf, Germany}
\footnotetext[20]{Now at Harvard University, Cambridge, MA 02138, U.S.A.}
\footnotetext[21]{Now at D\'epartement de Physique, Facult\'e des Sciences de Tunis, 1060 Le Belv\'ed\`ere, Tunisia.}
\footnotetext[22]{Supported by the US Department of Energy,
grant DE-FG03-92ER40689.}
\footnotetext[23]{Now at Department of Physics, Ohio State University, Columbus, OH 43210-1106, U.S.A.}
\footnotetext[24]{Also at Dipartimento di Fisica e Tecnologie Relative, Universit\`a di Palermo, Palermo, Italy.}
\footnotetext[25]{Now at CERN, 1211 Geneva 23, Switzerland.}
\footnotetext[26]{Now at ISN, Institut des Sciences Nucl\'eaires, 53 Av. des Martyrs, 38026 Grenoble, France.}
\footnotetext[27]{Now at Universit\`a degli Studi di Roma Tor Vergata, Dipartimento di Fisica, 00133 Roma, Italy.}
%
\setlength{\parskip}{\saveparskip}
\setlength{\textheight}{\savetextheight}
\setlength{\topmargin}{\savetopmargin}
\setlength{\textwidth}{\savetextwidth}
\setlength{\oddsidemargin}{\saveoddsidemargin}
\setlength{\topsep}{\savetopsep}
\normalsize
\newpage
\pagestyle{plain}
\setcounter{page}{1}

\pagestyle{plain}
\setcounter{page}{1}
\pagenumbering{arabic}
\normalsize
\setlength{\textheight}{23cm}
\setlength{\textwidth}{17cm}
\unitlength 1mm
\setlength{\topmargin}{-1cm}
\setlength{\oddsidemargin}{-0.5cm}
\section{Introduction}
\label{Introduction}

Searches for neutral Higgs bosons of the Standard Model, its minimal
supersymmetric extension (\MSSM), and extensions allowing Higgs boson
decays into invisible final states were performed using data collected
by the ALEPH detector at LEP during 1999.  The data sample was taken
at four centre-of-mass energies, 191.6, 195.5, 199.5, and 201.6\,\G\
at which 28.9, 79.9, 86.3, and 41.9\,\invpb\ of data were collected,
respectively.  The total data sample corresponding to an integrated
luminosity of $237\,\invpb$ was analysed to search for topologies
arising from the $\ee\to\h\Z$ Higgsstrahlung process supplemented by
\W\ and \Z\ gauge boson fusion, and from the $\ee\to\h\A$ associated
pair-production process of the
\MSSM.  The production cross section of the Higgsstrahlung process in
the \MSSM\ is reduced by a factor \sba, where \tanb\ is the ratio of
the vacuum expectation values of the two Higgs doublets and $\alpha$
is the mixing angle in the CP-even Higgs sector. The \h\A\ production
cross section is proportional to \cba.  For an invisibly decaying
Higgs boson, the observable rate of the Higgsstrahlung
process can be expressed as $\xi^2
\sigma_{\mathrm{SM}}(\epem\to\h\Z)$, where $\xi^2$ is the product of
the branching ratio to invisible decays and a model-dependent factor
which reduces the cross section with respect to that in the Standard
Model.

Searches for neutral Higgs bosons with the ALEPH detector have already
been carried out up to a centre-of-mass energy of
188.6\,\G~\cite{189paper, 189inv}; no evidence of a signal was found.
A lower limit at $95\%$ confidence level (CL) was set at 92.9\,\Gcs\
on the Standard Model Higgs boson mass. In the \MSSM\, for the
benchmark parameter set with maximal stop mixing~\cite{PhysatLEP2},
lower limits of 82.5\,\Gcs\ and 82.6\,\Gcs\ were derived on the masses
of the \h\ and \A\ Higgs bosons, respectively.  For an invisibly
decaying Higgs boson, a lower limit of $95.4\,\Gcs$ was set, for a 
production cross section equal to that in the Standard Model.
Similar searches have been performed by the other LEP 
experiments~\cite{otherLEP}.

The higher centre-of-mass energies and integrated luminosity in the
1999 data substantially increase the experimental sensitivity for the
detection of Higgs bosons with respect to previous results.
Nevertheless, the background processes are the same as
those described in Refs.~\cite{189paper, 189inv}.
The theoretical framework, the event selections, the study of
systematic uncertainties, and the result extraction are therefore very
similar to those previously described.  The differences mainly consist
of reoptimization of event selections, the introduction of new
discriminating variables, and improvements related to the simulation
of signal processes.

\section{ALEPH detector}
\label{detector} 

The components of the \ALEPH\ detector that are most relevant for the
analyses presented here are summarized in this section.  A more
detailed description of the detector can be found in
Ref.~\cite{ALEPHDET} and its performance in Ref.~\cite{ALEPHPERF}.

Three coaxial tracking devices are located inside a solenoidal superconducting
coil which produces an axial magnetic field of 1.5\,T.  The vertex
detector (VDET)~\cite{ALEPHVDET} consists of two cylindrical layers of
silicon wafers situated at average radii of 6.3 and 11.0\,cm.  Charged
particles with a polar angle in the range $\left|\cos\theta\right| < $
0.88 (0.95) traverse at least two (one) VDET layers.  The VDET is
surrounded by an inner tracking wire chamber (ITC) 
and by a large time projection chamber (TPC), which measures up to 21
three-dimensional coordinates per charged particle between radii of 31
and 180\,cm.
The tracking achieves a momentum resolution
$\sigma(\pt)/\pt$ of \mbox{$6 \times 10^{-4}~\pt \oplus 0.005$}, with
$\pt$ in \Gc.  The resolution on the three-dimensional impact
parameter of tracks can be parametrized as 
$(34 +70/p)(1+1.6\cos^{4}\theta)\,\mu$m, with $p$~in \Gc.

The electromagnetic calorimeter (ECAL) is also situated inside the
coil.  It is segmented into projective towers of typically
$0.9^{\circ} \times 0.9^{\circ}$, which allows electrons and photons
to be identified within jets.  Luminosity calorimeters of similar
construction to ECAL are installed between the endcaps and the beam
pipe and are treated as an extension of the ECAL.  A silicon-tungsten
sampling calorimeter extends the electromagnetic calorimeter coverage
down to 34\,mrad.  

Outside the coil, a hadron calorimeter (HCAL)
measures the hadronic energy, acts as a filter for the identification
of muons, and serves as a return yoke for the magnetic field.  The
outermost detectors are two double layers of muon chambers.

The measurements of charged particle tracks and of energy deposits in
the calorimeters, combined with the identification of photons,
electrons, and muons, are used to produce a list of charged and
neutral energy flow particles.  Hadron jets, formed by
clustering these particles, have an energy resolution of 
$\sigma(E) =(0.60\sqrt{E} + 0.6)(1+\cos^{2}\theta)$, where $E$ is the energy
in GeV and $\theta$ the polar angle of the jet.  The resolution on the jet
angles is approximately \mbox{20\,mrad} in both $\theta$ and $\phi$.

The tagging of b quark jets is accomplished by combining several
discriminating variables together using neural networks.  These
variables include an impact-parameter-based
probability~\cite{qipbtag}, a displaced vertex $\chi^2$~\cite{QVSRCH},
and the transverse momentum of identified leptons with respect to the
jet axis~\cite{duccio}.  Three jet-shape quantities supplement these
variables.
The resulting six variable neural network b-tag is described in
Ref.~\cite{161paper}; a more specialized four variable b-tag used for
the $\ee\to\h\A\to\bb\bb$ final state is described in
Ref.~\cite{189paper}.

\section{Update of the searches}\label{update}
\subsection{Signal and background simulation}
\label{mc} 

For each of the four centre-of-mass energies, fully simulated samples
of signal and background events were generated.  The \HZHA\
program~\cite{HZHA}, used to generate the signal events, was also used
to compute the signal cross sections, the Higgs boson decay branching
fractions, and the radiative corrections to the Higgs boson masses in
the \MSSM.  The most recent version contains, in particular, the
latest refinements on these radiative corrections~\cite{PhysatLEP2,
subhpole2, Benchmarks} and a more complete simulation of the
\h\e$^{+}$e$^{-}$ and
\h$\nu_{\mathrm{e}} {\bar\nu_{\mathrm{e}}}$ final states which now
includes the interference between the boson fusion and the
Higgsstrahlung processes~\cite{Zerwasetal}.

The $\ee\to\qq(\gamma)$\ background was generated with
\KORALZ~\cite{KORALZ}, instead of
\PYTH~\cite{PYTH}, to benefit from a more accurate treatment of multiple
initial state photon radiation.  The $\ee\to\W\e\nu$ events, generated
with \PYTH, were compared to events generated with \linebreak
\GRACE~\cite{GRACE4F}, and the differences were taken into
account in the systematic uncertainties.
The other Standard Model four-fermion processes were simulated in a
manner similar to that described in Refs.~\cite{189paper, 189inv}.
The sizes of the simulated signal and background samples correspond to
at least 50 times the collected luminosity used in the analyses.

\subsection{Event selections}
Analyses had already been developed~\cite{189paper, 189inv} for
most final states relevant for the searches for neutral Higgs bosons.  
For the visible $\ee\to\h\Z$ search, these final
states are the leptonic final state (\h$\ell^{+}\ell^{-}$ where $\ell$
denotes an electron or a muon), the missing energy final state
(\h\nunu), the four-jet final state (\h\qqbar), and the tau 
final states (\h\tptm\ and \h\to\tptm, \Z\to\qqbar).  For the
$\ee\to\h\A$ search, they are the four-b final state (\bbbar\bbbar)
and the tau final state (\bbbar\tptm).  For the invisible
$\ee\to\h\Z$ search, they are the acoplanar lepton pair final state
(\h\lplm) and the acoplanar jet pair final state (\h\qqbar).  
As in Ref.~\cite{189paper}, the visible Higgs boson search was
conducted in two alternative analysis streams, a neural-network-based
stream (NN) and a cut-based stream (cut), although both streams
share the same leptonic and tau-final-state selections. 
This section highlights changes with respect to Refs.~\cite{189paper,
189inv}.
Most of the \h\Z\ and \h\A\ selections were optimized
for Higgs boson masses of 107 and 90\,\Gcs, respectively, {\it i.e.}, close
to the search sensitivity in both cases.

A reoptimization of the selection criteria and retraining of the
neural networks in all channels were performed to account for the
higher centre-of-mass energies and integrated luminosity.  In some
cases, several new discriminating variables were adopted and
refinements to analysis techniques implemented, as described below.

\begin{figure}[t]
\begin{center}
\begin{picture}(180,80)
\put(40,0){\epsfxsize=85mm\epsfbox{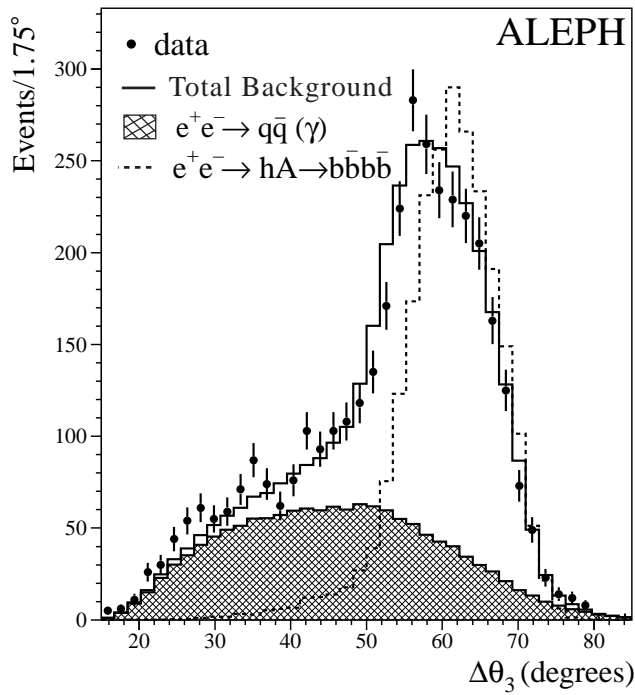}}
\end{picture}
\caption{\capstyl Distribution, at preselection level, of 
the minimum value of the energy-weighted angular dispersion
$\Delta\theta_3$ in the four-b channel for data (dots with error
bars), simulated background (solid histogram), and the signal for $\mh
= \mA = 90$\,\Gcs~(dashed histogram).  The signal histogram has an
arbitrary normalization.}
\label{ang3}
\end{center}
\end{figure}

\begin{figure}[ht]
\begin{center}
\begin{picture}(180,80)
\put(40,0){\epsfxsize85mm\epsfbox{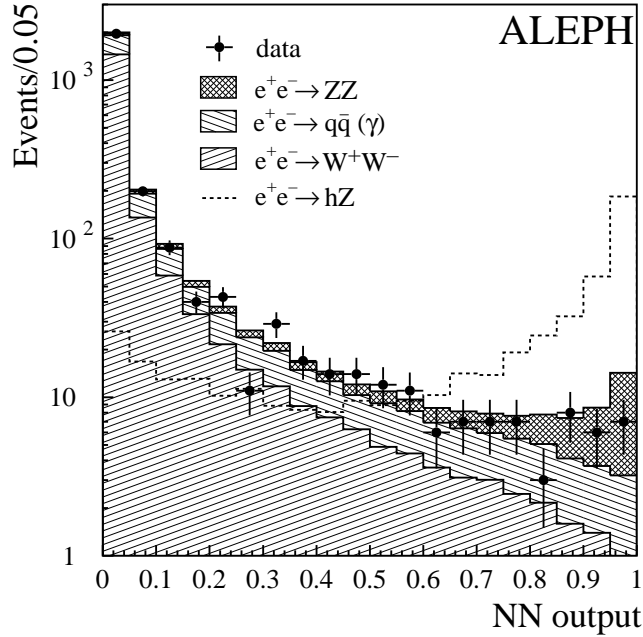}}
\end{picture}
\caption{\capstyl
Distributions of the neural network output used to select
four-jet \h\Z\ candidates for data (dots with error bars), simulated
background (solid histograms), and simulated Higgs signal with an
arbitrary normalization for $\mh =
107$\,\Gcs\ (dashed histogram), after
preselection criteria have been applied.
\label{fig:4jet}}
\end{center}
\end{figure}

\begin{itemize}
\item In the missing energy channel, the 
reconstructed Higgs boson mass was
removed from the variables of the single neural network selection
(referred to as ``A'') in order to increase the selection efficiency
for low Higgs boson masses.
The structure of the three-neural-network selection (referred to as
``B'') remains unchanged.

\item In the tau final state selection, the overlap with
leptonic-final-state events ($\h\epem$ or $\h\mpmm$) was reduced by
requiring that the measured invariant mass of the two tau jets be
smaller than 75\,\Gcs\ when a decay particle of at least one of
the two taus is identified as an electron or a muon.

\newcounter{bean}
\item In the four-jet channels, the following improvements were made.
\begin{list}
{\roman{bean}) }{\usecounter{bean}\setlength{\rightmargin}{\leftmargin}}
\item A four-constraint fit, identical to that used in Ref.~\cite{4cfit},
 was adopted in all selections, instead of a simple energy
 rescaling. It allows the jet angles to be fitted as well as the jet
 energies, thus improving the dijet mass resolution, particularly near
 the kinematic threshold.

\item In the four-b channel, a new variable was introduced
      to characterize the unbalanced topology of three jets recoiling
      against one.  For each of the possible three-jet combinations, the
      energy-weighted angular dispersion is computed as $$
\Delta_{ijk} \equiv
\left( 
\sum_{l \in {\rm jet}\, i,j,k} \theta^2_{l,\vec{u}} E_l 
\Bigg/
\sum_{l \in {\rm jet}\, i,j,k} E_l 
\right)^{1/2}
      $$ where the indices $i$, $j$, and $k$ refer
      to jets, $\theta_{l,\vec{u}}$ is the angle between the momentum
      direction of the object $l$ contained in jet $i,j,k$ and the
      vectorial sum $\vec{u}$ of the
      three jet momenta, and $E_l$ is the energy of object
      $l$.  The variable $\Delta\theta_3$ is defined as the minimum
      value of this dispersion for all four possible three-jet
      systems.
      The distribution of this quantity is shown in Fig.~\ref{ang3}.
      The cut $\Delta\theta_3>50^{\circ}$ reduces the background from
      events with gluon splitting \epem\to \bb (g\to\bb).

\item In the neural network selection, the
      former 17 variables were supplemented with the decay angles
      relative to the direction of flight 
      of the Z and Higgs boson candidates, $\alpha_{12}$
      and $\alpha_{34}$.
The output of the neural network for the four-jet final state
is shown in
Fig.~\ref{fig:4jet}. 

\end{list}
\end{itemize}

\vspace{1.0cm}

\vspace{-1cm}
\subsection{Systematic uncertainties}
\label{syste}
Systematic uncertainties caused by inaccuracies of the simulation were
evaluated as in Refs.~[1,~2].  Whenever possible, they
were extracted from 3.5\,\invpb\ of data taken at the \Z\ peak during
the same year.  A discrepancy between data and simulation, affecting
the acceptance for signal and for most of the background processes,
was observed in the impact-parameter-based b tagging quantities.  To
correct for this effect, a smearing of the track parameters was
performed on the simulated events to bring them in better agreement
with the data.  Half of the correction was conservatively taken as a
systematic uncertainty for all channels.
The systematic uncertainties on signal efficiencies are
typically $5\%$, and those on the background yields are summarized 
in Table~\ref{bigtable}.

\begin{table}[t]
\caption{\capstyl The numbers of signal and background events expected, and 
the numbers of candidate events observed in the data for each channel.
The signal expectation is computed for the \h\Z\ process with a
Higgs boson mass
of 107\,\Gcs, and that for the \h\A\ process with a common
Higgs boson mass of 90\,\Gcs. The background yield is subdivided into
\Z\Z\ (including Zee and \Z\nunu), \W\W\ (including $\W\e\nu$), 
and \ffbar\ (including $\gamma\gamma\rightarrow\ffbar$).  The \ffbar\
composition is entirely $\qq$ for the visible Higgs boson and hadronic
invisible Higgs boson decay channels, and $\lplm$ and
$\gamma\gamma~\to\lplm$ for the leptonic invisible Higgs boson decay
channel.  Systematic uncertainties are given for each background.
\label{bigtable}}
\begin{center}
\begin{tabular}{|c|l|r|r|r@{$\pm$}lr@{$\pm$}lr@{$\pm$}l|r@{$\pm$}l|c|}
\hline\hline
\multicolumn{2}{|c|}{Analyses} &
\multicolumn{2}{c|}{Signal} &
\multicolumn{8}{c|}{Background} & Events \\
\multicolumn{2}{|c|}{      } &
\multicolumn{2}{c|}{expected} &
\multicolumn{8}{c|}{expected} & Observed \\

\cline{3-12}
 \multicolumn{2}{|c}{ } & \multicolumn{1}{|c|}{\h\Z} & \multicolumn{1}{c|}{\h\A} &
     \multicolumn{2}{c}{\Z\Z} & \multicolumn{2}{c}{\W\W} & \multicolumn{2}{c|}{\rule{0cm}{0.45cm}\ffbar} &
     \multicolumn{2}{c|}{Total} &  \\
\hline
\multicolumn{2}{|c|}{\h\lplm} &
  0.8 &                  
\multicolumn{1}{|c|}{--} &      
 25.4 & 0.3  &                  
  1.9 & 0.1  &                  
  1.2 & 0.2  &                  
 28.5 & 0.4  &                  
 26   \\\hline\hline
\multicolumn{2}{|c|}{\h\nunu\ (Cut)} &
 2.6 &
\multicolumn{1}{|c|}{--} &
11.1  & 1.1 &
10.3  & 3.1 &
3.7   & 2.2 &
25.1  & 4.0 &
23 \\\hline\hline
\h\nunu & A \& B & 
1.1   &
\multicolumn{1}{|c|}{--} &
2.5  & 0.1 &
0.2  & 0.0 &
0.5  & 0.1 &
3.2  & 0.1 &
3 \\\cline{2-13}
 (NN)      & A only&
0.6   &
\multicolumn{1}{|c|}{--} &
4.2   & 0.1 &
1.0   & 0.1 &
0.6   & 0.1 &
5.8   & 0.2 &
3 \\\cline{2-13}
      & B only&
0.2  &
\multicolumn{1}{|c|}{--} &
0.5   & 0.0 &
0.1  & 0.0 &
1.4  & 0.2 &
2.0   & 0.2 &
1 \\\hline\hline
\h\qq & \h\Z\ \& \h\A & 
1.5   & 
2.4  &
5.0  & 0.4 &
0.3   & 0.1 &
2.5  & 0.4 &
7.8  & 0.8 &
3 \\\cline{2-13}
 (Cut)      & \h\Z\ only &
2.2   &
0.3  & 
10.8   & 0.6 &
2.9  & 0.3 &
5.9   & 0.5 &
19.7   & 1.0 &
16 \\\cline{2-13}
      & \h\A\ only &
0.3   & 
1.2  &
2.3   & 0.2 &
1.2  & 0.2 &
5.3   & 0.6 &
8.8   & 0.8 &
8 \\\hline\hline
\h\qq & 2b & 
4.1   & 
1.0  &
19.7  & 1.0 &
7.7   & 0.6 &
12.3  & 1.5 &
39.7  & 1.9 &
27 \\
\cline{2-13}
 (NN)      & 4b &
1.1   &
2.2  &
3.9   & 0.3 &
0.3  & 0.1 &
2.5   & 0.7 &
6.7   & 0.8 &
3 \\\hline\hline
\rule{0pt}{4.3mm} \bbbar\tptm & \h\Z\ \& \h\A &
0.2   &
0.4   &
1.8   & 0.2 &
0.4   & 0.1 &
0.1   & 0.1 &
2.3   & 0.2 &
0 \\\cline{2-13}
\&          & \h\Z\ only &
0.3   & 
0.1   &
4.3   & 0.2 &
4.6   & 0.2 &
0.8   & 0.2 &
9.7   & 0.3 &
13 \\\cline{2-13}
\tptm\qq    & \h\A\ only &
0.0   & 
0.0   &
0.2   & 0.0 &
0.1   & 0.0 &
0.1   & 0.1 &
0.4   & 0.1 &
2 \\\hline \hline
\multicolumn{2}{|c|}{invisible \h\lplm} &
  0.3 &                  
\multicolumn{1}{|c|}{--} &      
  3.4 & 0.4  &                  
  3.7 & 0.5  &                  
  1.4 & 0.2  &                  
  8.5 & 0.7  &                  
 8   \\\hline\hline
\multicolumn{2}{|c|}{invisible \h\qqbar} &
  3.3 &                  
\multicolumn{1}{|c|}{--} &      
  31.8 & 0.8  &                 
  17.4 & 3.5  &                 
  3.1 & 1.2  &                  
  52.3 & 3.8  &                 
 50   \\\hline\hline
\end{tabular}
\end{center}
\end{table}

\section{Combination and results}
\label{results}
\subsection{Selection overlaps}
\label{strategy} 

Events selected by more than one analysis were treated with a
well-defined prescription to remove overlaps.  This procedure prevents
any given event from affecting the confidence level calculations more
than once.  The potential overlaps of the searches for different
topologies arising from the Higgsstrahlung process were suppressed by
assigning a precedence order to the four selections, and by rejecting
any candidate event already selected by a higher precedence
search. The order of the precedence was chosen to be 1)~the missing
energy final state; 2)~the tau final states; 3)~the leptonic final
state; and 4)~the four-jet final state.  In the data, no event was
affected by this procedure.

The significant overlaps of the hZ and hA searches for tau final
states as well as for four-jet final states in the cut stream were
treated in a manner similar to that used in Ref.~\cite{189paper} for
the missing energy channel.  This method divides the analyses into
three statistically independent branches: one branch containing the
events selected only by the hZ search; a second branch containing the
events selected only by the hA search; and a third branch containing
the events selected simultaneously by both searches.  The use of
different reconstructed Higgs boson mass definitions in the hZ
searches (with the Z mass constraint) and the hA searches (with an
equal mass constraint) introduces an additional difficulty, solved in
the following manner. The hZ (hA) mass definition is used for the
hZ(hA)-only branch. For the third branch, the mass definition depends
on the MSSM parameter set and is chosen such that it maximizes the
combined expected $95\%$ CL sensitivity. Typically, the hZ mass
definition is chosen for small \tanb\ values, where the hZ cross
section dominates, and the hA mass definition is chosen for large
\tanb\ values.

The three-branch subdivision in the NN stream four-jet final state
was replaced by a simpler and similarly sensitive two-branch treatment:
the first branch (4b) containing the events selected by the four-b
final state search, and the second branch (2b) containing the events
not selected by the 4b search.  

\vspace{-0.4cm}
\subsection{Selection results}
\label{combin} 

\begin{figure}
\begin{center}
\begin{picture}(180,80)
\put(0,0){\epsfxsize80mm\epsfbox{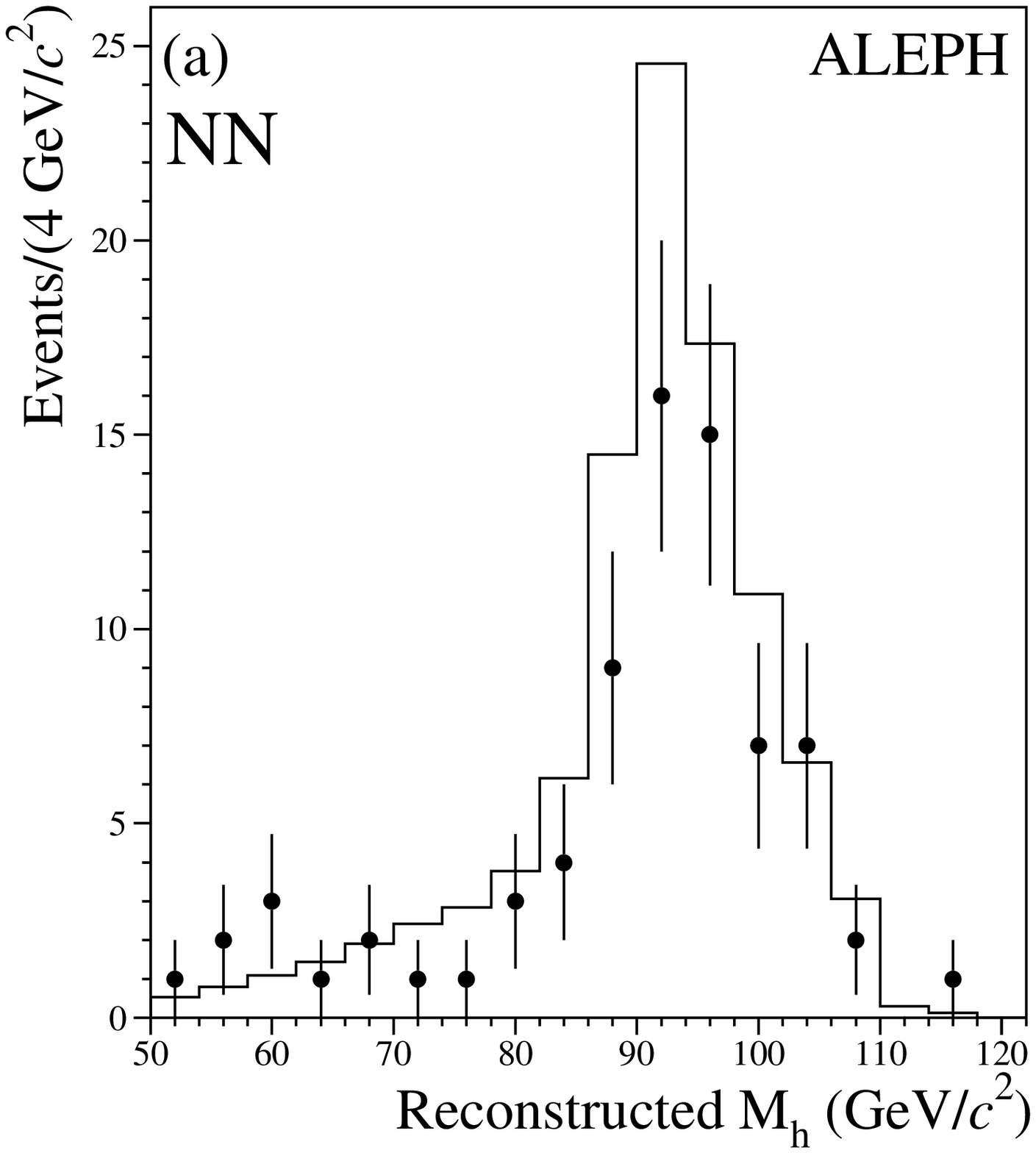}}
\put(85,0){\epsfxsize81.5mm\epsfbox{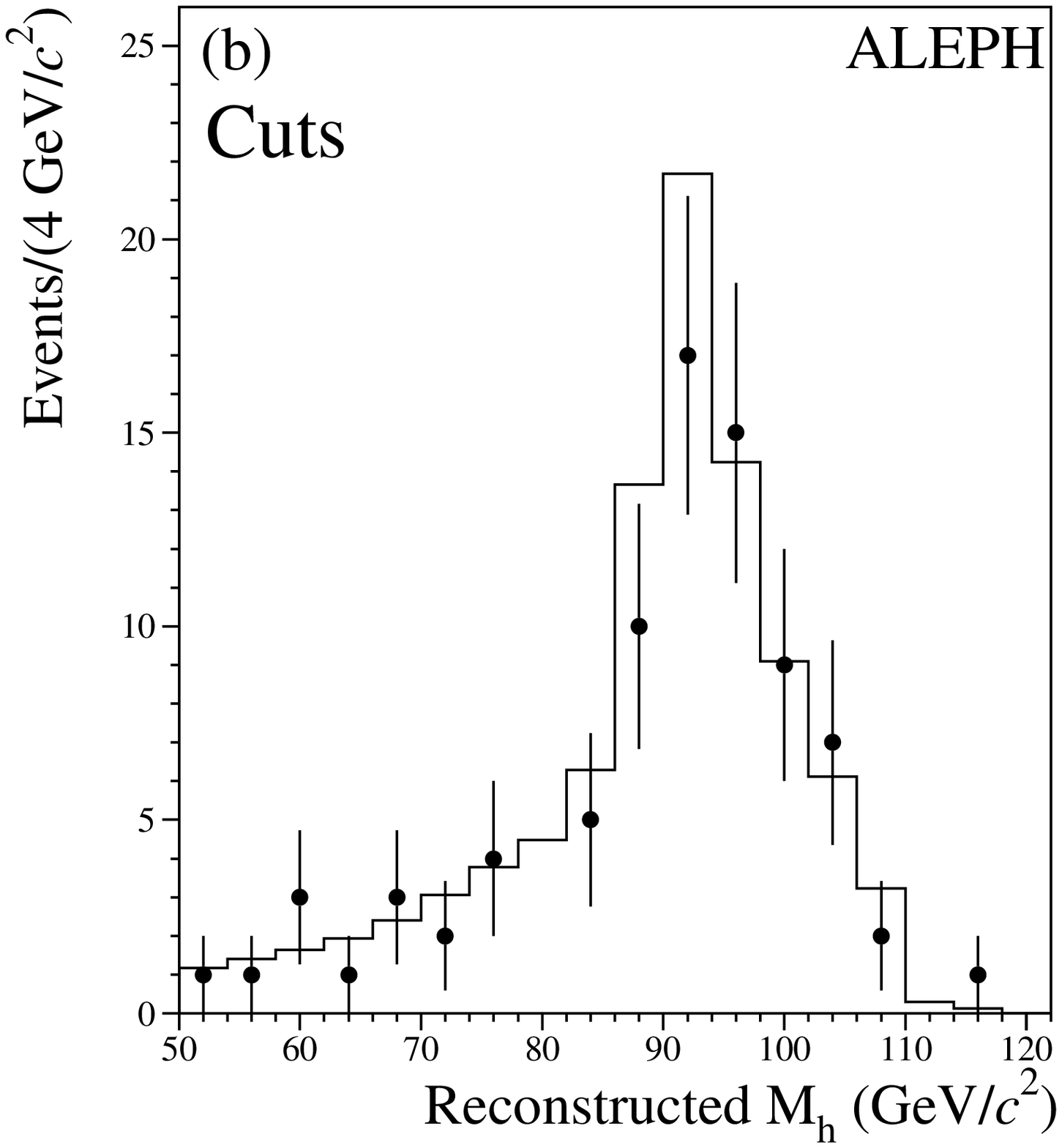}}
\end{picture}
\caption{\capstyl Distributions of the reconstructed Higgs boson mass $M_{\mathrm{h}}$ for the  
data (dots with error bars) selected in the \h\Z\ searches by (a) the NN
stream, and (b) the cut stream.
The histograms show the Standard Model background expectation. 
\label{mass_plot}} 
\end{center}
\end{figure}

\begin{figure}[ht]
\begin{center}
\mbox{\epsfxsize=.5\hsize\epsfbox{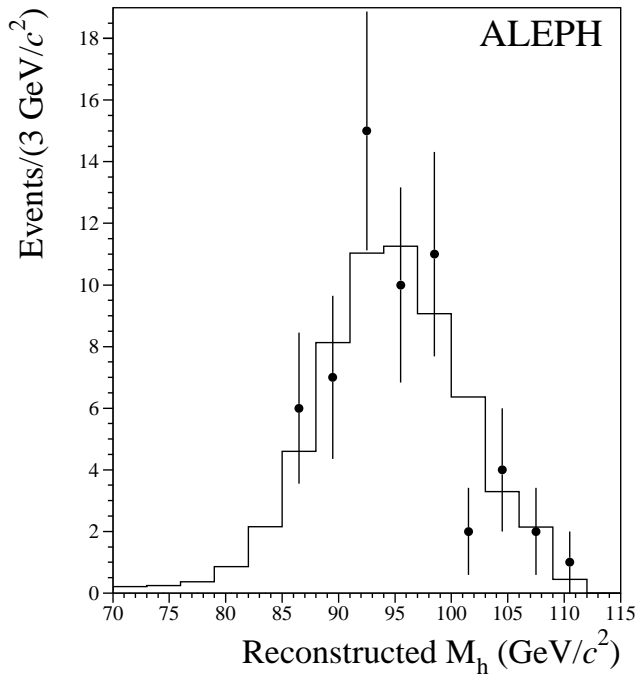}}
\caption{\capstyl Distribution of the reconstructed Higgs boson 
mass $M_{\mathrm{h}}$ for the data (dots with error bars) and the
expected background (histogram) in the invisibly decaying Higgs boson
searches for the acoplanar lepton and the acoplanar jet final states.
\label{fig:invmass}} 
\end{center}
\end{figure}

For each analysis, the numbers of signal and background events
expected, and the observed number of events are summarized in
Table~\ref{bigtable}.  For the Standard Model Higgs boson searches, a
total of 76 events are selected in the data with the NN stream, while
97.8 events are expected from Standard Model processes.
Figure~\ref{mass_plot} shows the reconstructed Higgs boson mass
distributions for the data and the background expectations for the
\h\Z\ selections.  The \h\A\ selections yield 5 events in the data
with the NN stream, while 9.4 events are expected from Standard Model
processes.  In the search for invisibly decaying Higgs bosons, a total
of 58 candidate events are selected with 60.8 events expected from
Standard Model processes; the distributions of their reconstructed
Higgs boson mass are shown in Fig.~\ref{fig:invmass}.

\label{deficit}

Fewer events were selected in the data than were expected. This
deficit is mostly due to the four-jet analysis of the NN stream where
46.4 events were expected, but only 30 events were selected in the
data.  In the four-jet cut analysis 27 events are selected in data
with 36.3 expected.  As can be seen in Fig.~\ref{mass_plot}, the
deficit of events is mostly near the mass region dominated by
\ee\to\ZZ\ events.

At the preselection level, the background expectation is dominated by
four-jet \linebreak 
$\ee\to\WW$ events which are kinematically similar
to signal events, and no significant discrepancies were found between
data and Monte Carlo simulation for all relevant kinematic variables.
Figure~\ref{fig:4jet} shows the distribution of the neural network
output distribution at the preselection level. Overall, the agreement
between the simulated expectation and the data is very good with 2572
events expected and 2523 seen in the data. The deficit is, however,
apparent in Fig.~\ref{fig:4jet} at high values of the neural network
output, which is dominated by b~tagged \ee\to\ZZ\ events.

As mentioned in Section~\ref{syste}, a slight disagreement was found
and corrected for in the b~tagging distributions. The corrections were
determined from \Z\ peak data taken during the same year.  These were
verified with $\ee\to\qqbar\gamma$ radiative return events and
semileptonic $\ee\to\WW\to\qq\ell\nu$ events collected at high
centre-of-mass energies.  The $\qqbar\gamma$ events provide a known
$\bbbar$ content while the semileptonic \WW\ events
provide a control sample with negligible b-jet content.  The impact of
these b-tagging corrections is propagated to the simulation, but the
effect is too small to account for the observed deficit of events.
The conclusion of these studies is that the cause of the deficit is
most likely due to a statistical fluctuation.

\subsection{Exclusion limits}
The $176.2\,\invpb$ of data collected in 1998 with a centre-of-mass
energy of $188.6\,\G$ were combined with the 1999 data to determine
the final results.  As was done in Refs.~\cite{189paper, 189inv},
confidence levels on the signal hypothesis were drawn using the Signal
Estimator method~\cite{SEpaper} rather than with the method used by
the other LEP experiments~\cite{ARead}, thus increasing the expected
mass limit by $\sim 0.5\,\Gcs$.  Expected confidence levels were
calculated as the median confidence level in the absence of signal.

The tau final state analyses and the cut analyses for the four-jet,
four-b, and missing energy topologies use only the reconstructed Higgs
boson mass as a discriminant in the confidence level calculations.
The other analyses use an additional discriminant variable: the neural
network output in the case of the 2b branch of the NN stream four-jet
analysis and a b-tag sensitive variable in the 4b branch as well as in
the \h\lplm\ final state analysis.

\begin{figure}
\begin{center}
\begin{picture}(180,80)
\put(0,0){\epsfxsize80mm\epsfbox{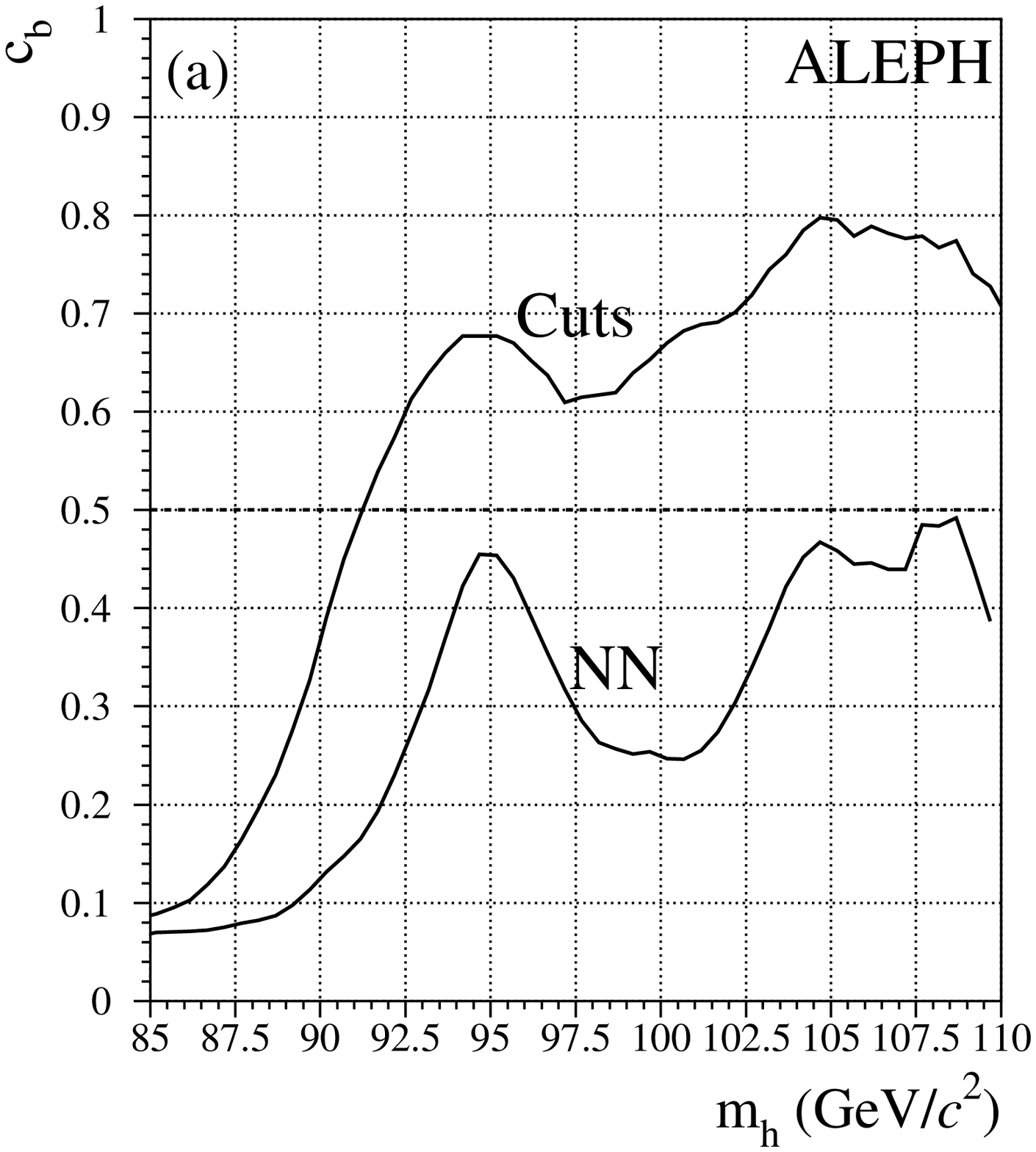}}
\put(85,0){\epsfxsize82mm\epsfbox{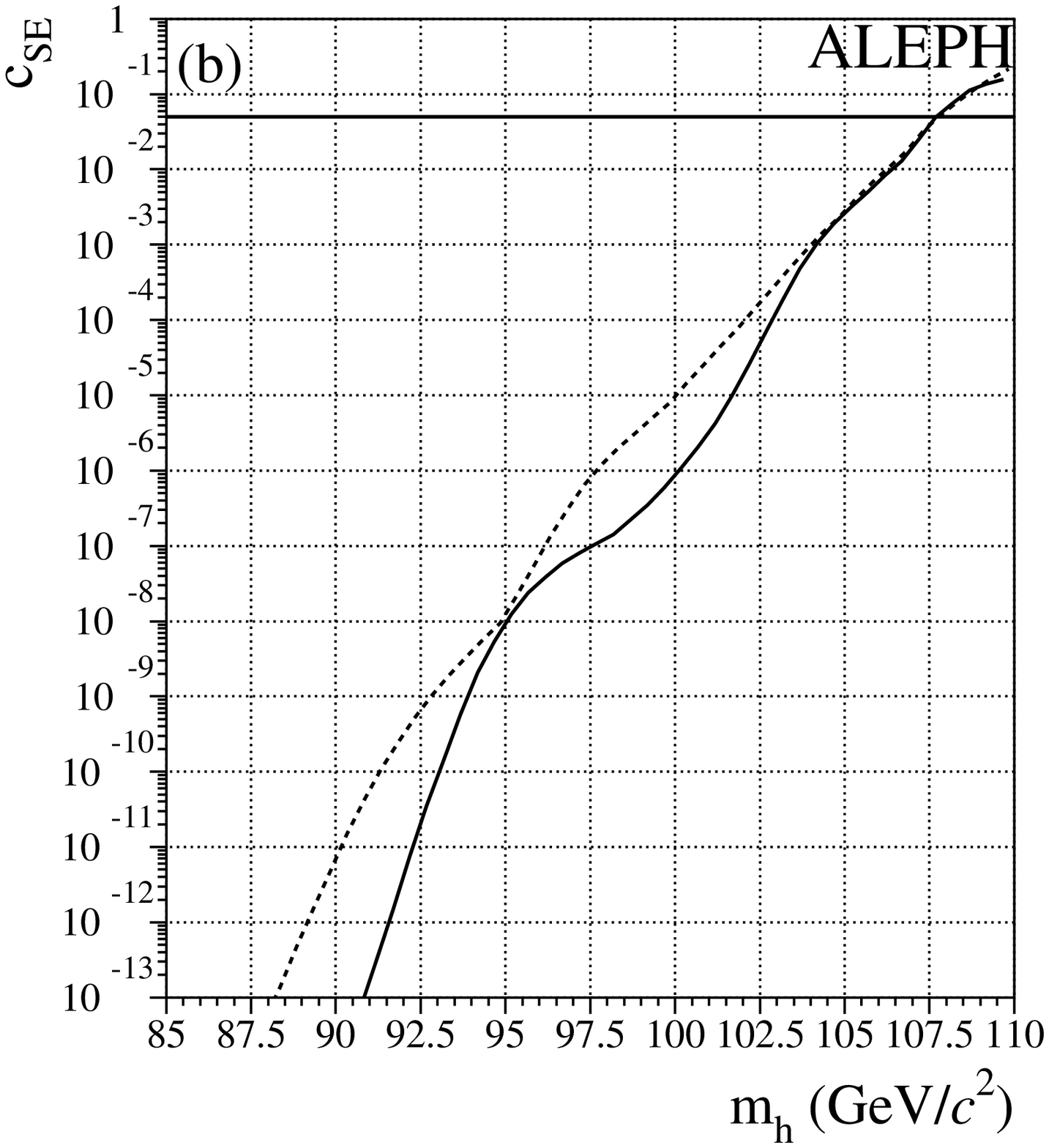}}
\end{picture}
\caption{\capstyl
Observed (solid) and expected (dashed) CL curves for (a) the
background hypothesis and (b) the signal hypothesis for the NN stream,
as functions of the hypothesized Standard Model Higgs boson mass
$m_{\mathrm{h}}$.  For the signal hypothesis (b), the intersections of
the horizontal line at $5\%$ with the curves define the observed and
expected $95\%$ CL lower limits on the Standard Model Higgs boson
mass.
\label{fig:nn_clevel}}
\end{center}

\vspace{0.15cm}
\begin{center}
\begin{picture}(180,80)
\put(18,80.5){\bf (a)}
\put(103,80.5){\bf (b)}
\put(0,0){\epsfxsize83mm\epsfbox{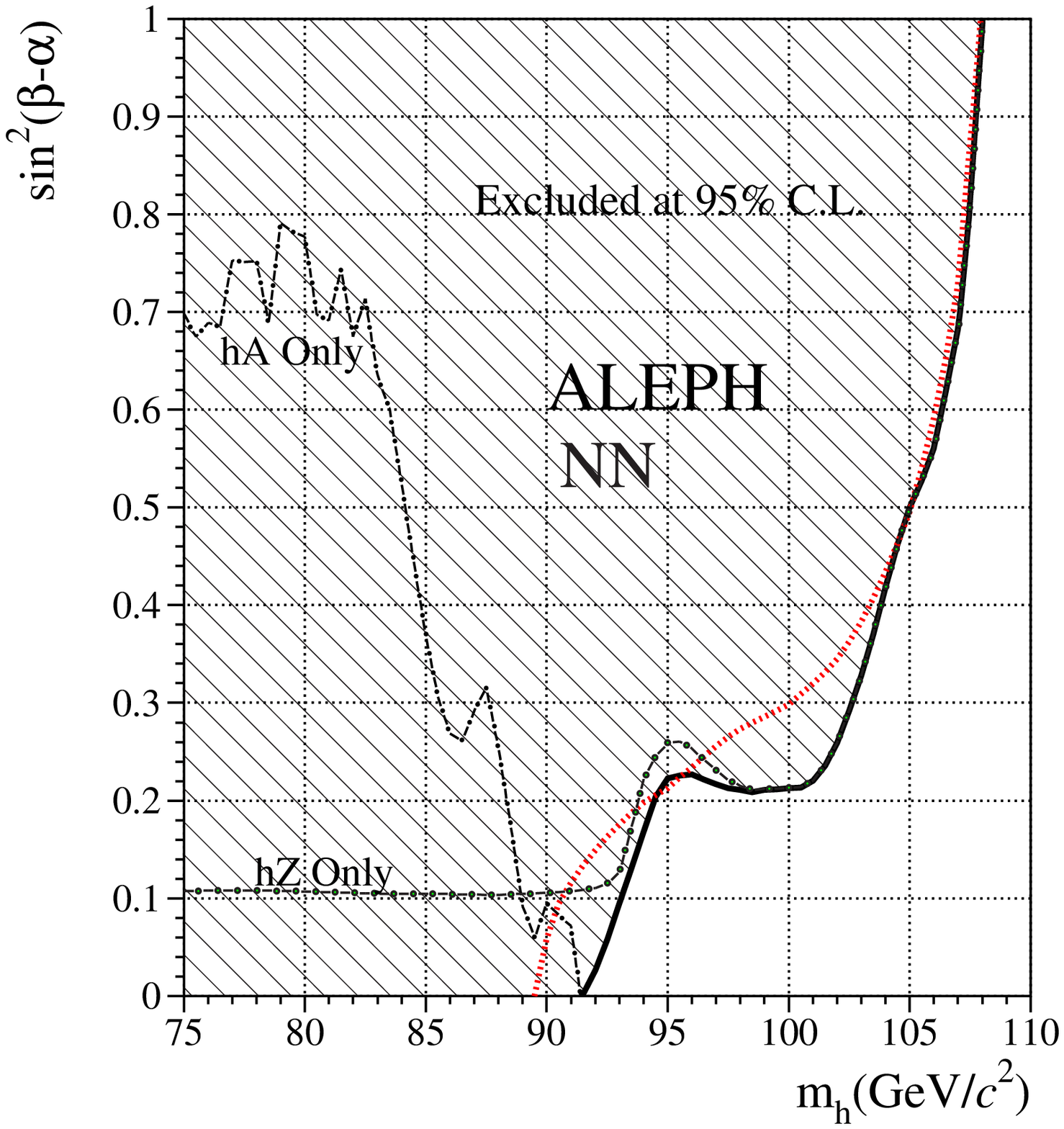}}
\put(85,0){\epsfxsize83mm\epsfbox{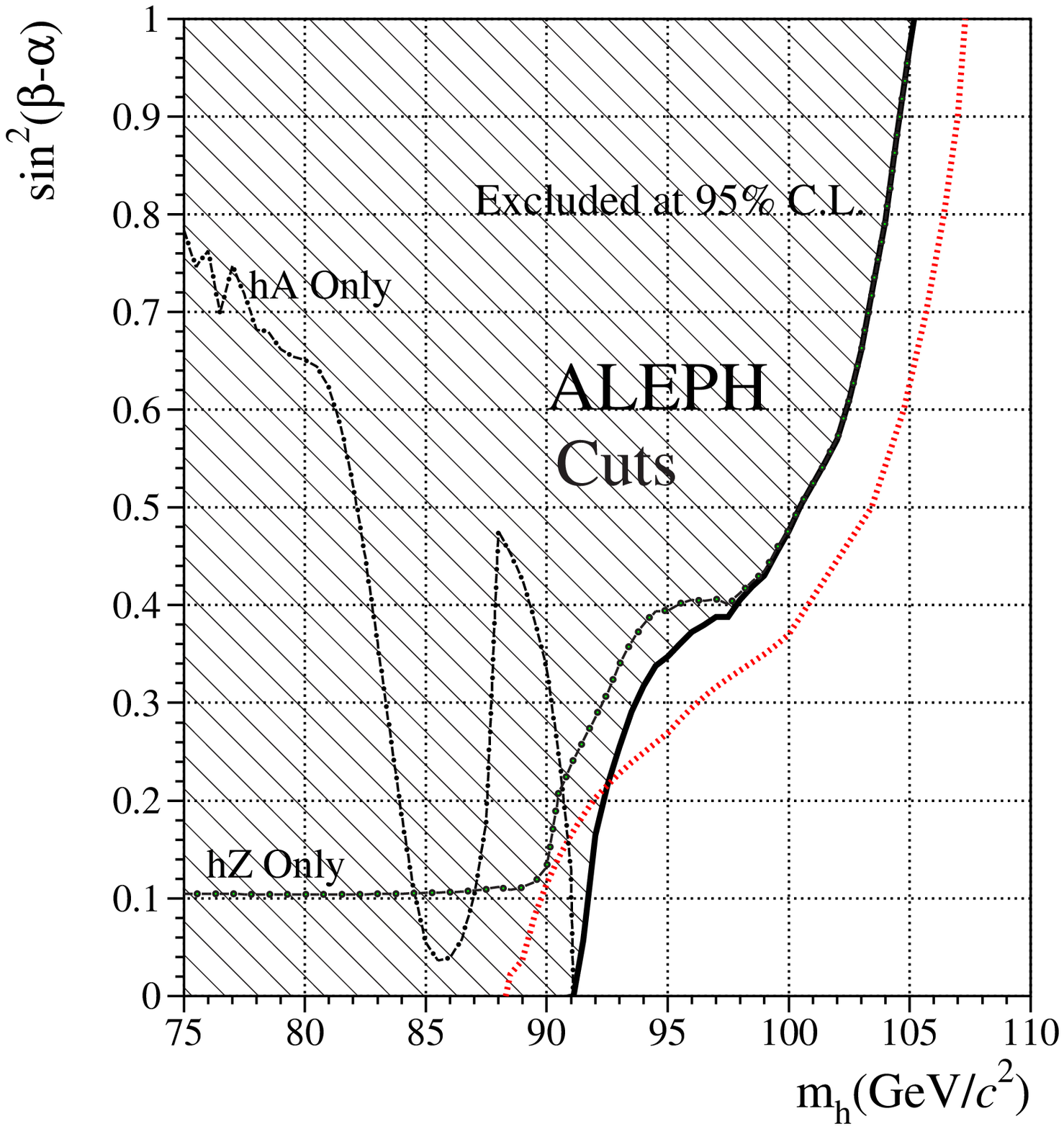}}
\end{picture}
\caption{\capstyl
The expected (dashed) and observed (solid) excluded regions at $95\%$
CL in the plane
$[m_\mathrm{h},\sin^{2}(\beta-\alpha)]$ for (a) the NN
stream and (b) the cut stream.
\label{fig:nn_mssm}}
\end{center}
\end{figure}

The expected and observed confidence levels are shown in
Fig.~\ref{fig:nn_clevel} for the Standard Model Higgs boson searches.
The low confidence levels for the background hypothesis at small Higgs
boson masses are a reflection of the observed deficit of events
discussed in Section~\ref{deficit}.
The lower limits at $95\%$ confidence level on the mass of the
Standard Model Higgs boson with systematic uncertainties taken into
account~\cite{cousin} are summarized in Table~\ref{nn_summarytable}.  Since 
the NN stream gives an expected Higgs boson mass lower limit
higher by about 0.5~\Gcs, it is used to set the final results.

\begin{figure}[h!]
\begin{center}
\begin{picture}(180,80)
\put(0,0){\epsfxsize80mm\epsfbox{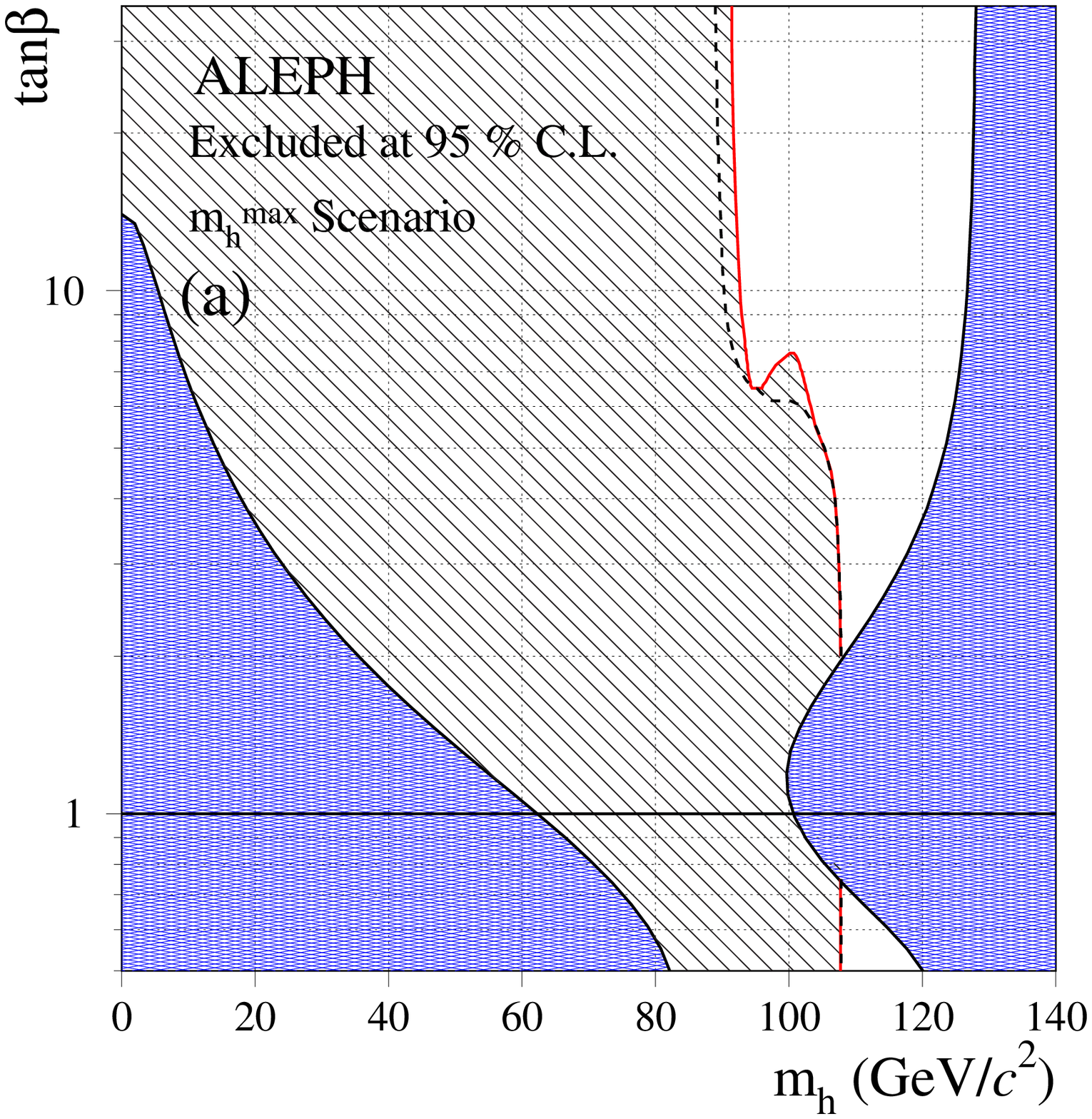}}
\put(85,0){\epsfxsize80mm\epsfbox{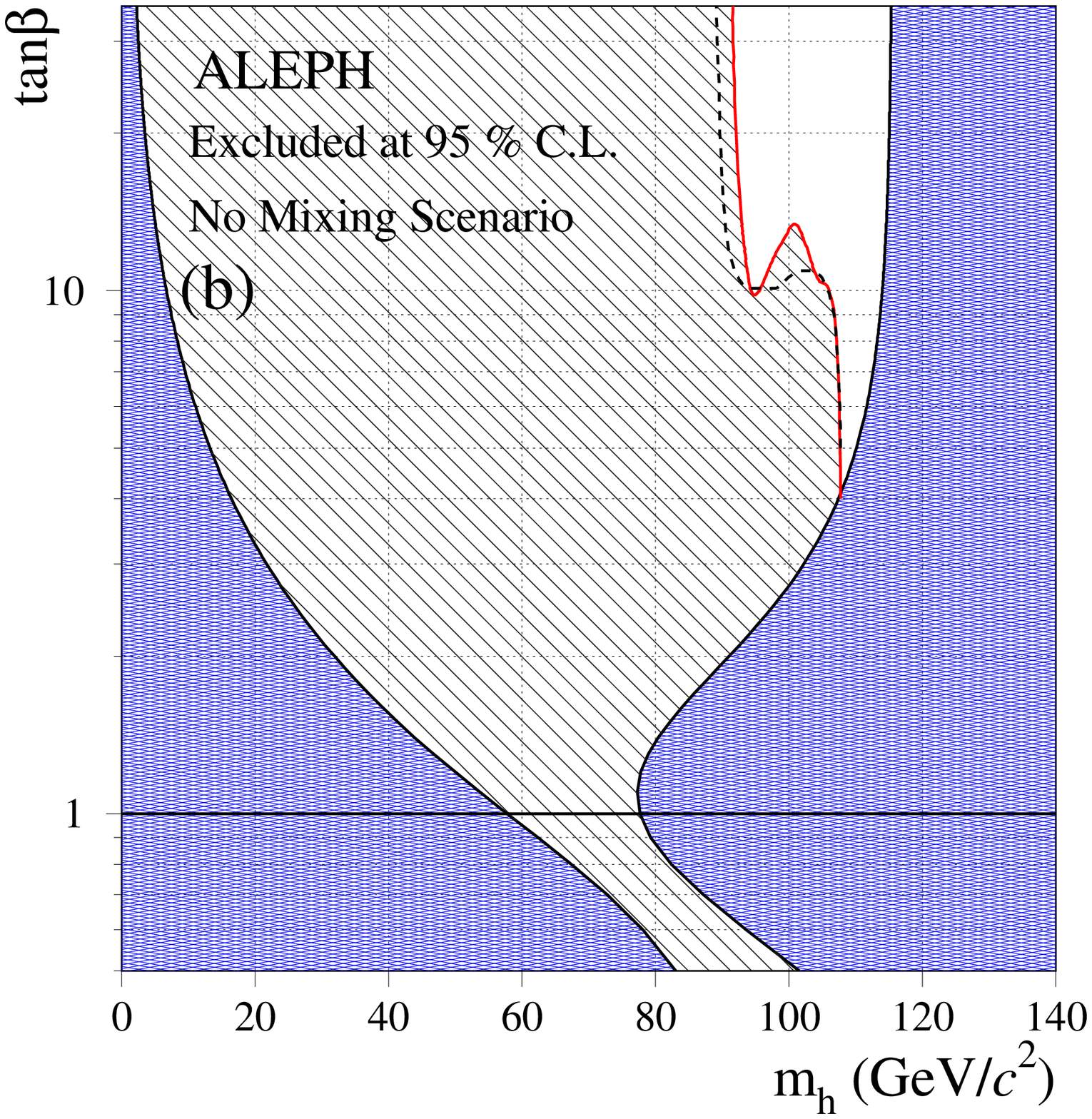}}
\end{picture}
\caption{\capstyl
The expected (dashed) and observed (solid) excluded regions at
$95\%$ CL for the NN stream in the [\mh, \tanb] plane of the \MSSM\ for
(a) the $m_{\mathrm{h}}^{\mathrm{max}}$ parameter set and (b) the no
mixing scenario.  The densely hatched regions are not allowed
theoretically.
\label{fig:nn_tanb}}
\end{center}
\vspace{0.75cm}
\begin{center}
\begin{picture}(180,80)
\put(37.5,0){\epsfxsize95mm\epsfbox{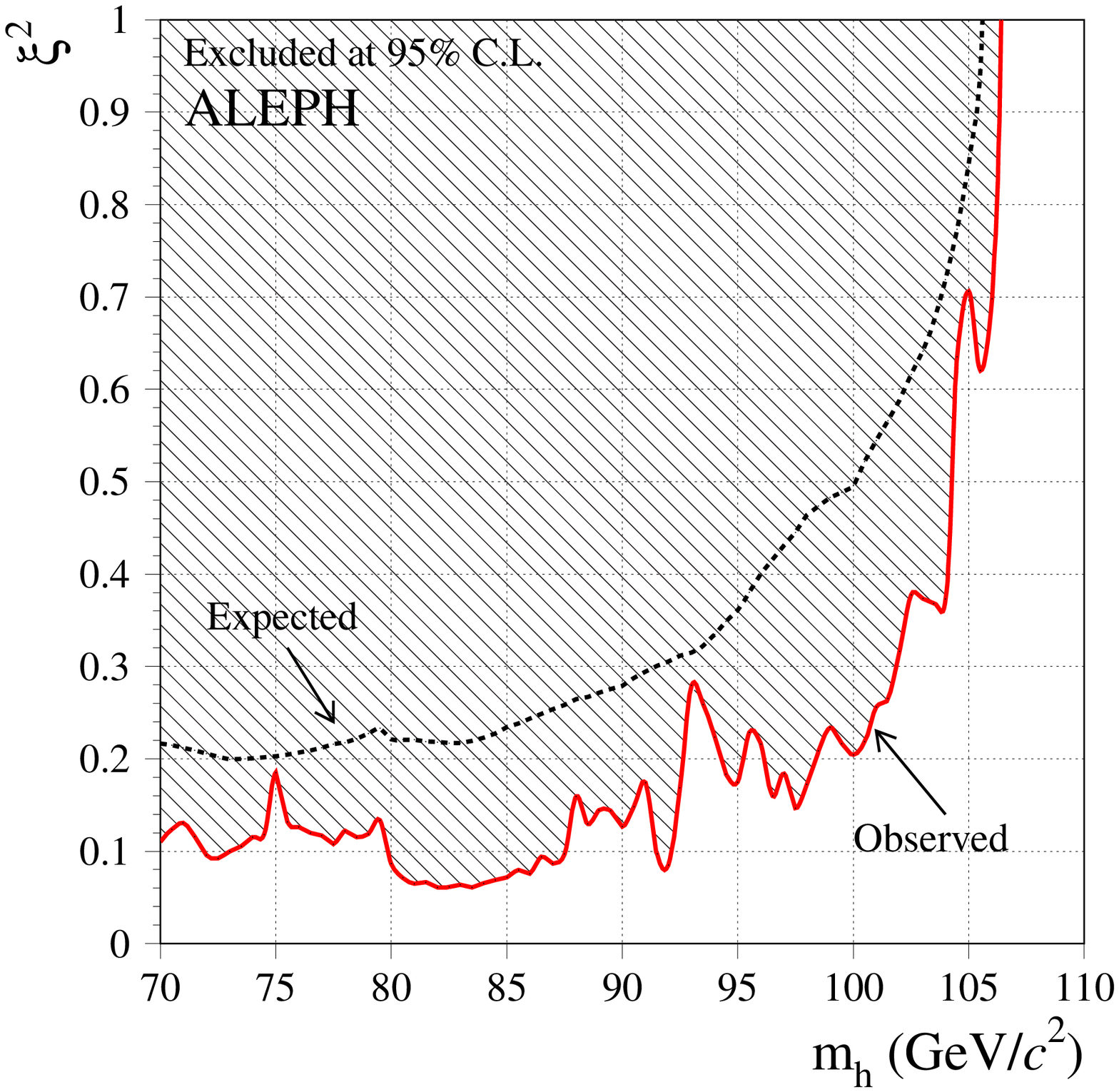}}
\end{picture}
\caption{\small Observed (solid) and expected (dashed) regions excluded at 
the $95\%$ CL in the (\mh,$\xi^2$) plane for an
invisibly decaying Higgs boson.
\label{plan}}
\end{center}
\end{figure}

Figure~\ref{fig:nn_mssm} shows the \h\Z, the \h\A, and their combined
Higgs boson mass exclusions as a function of \sinba.  These combined
exclusions are interpreted within two benchmark parameter sets, one
which maximizes the radiative corrections to the lighter CP-even Higgs
boson mass as a function of \tanb\ (referred to as the
$m_{\mathrm{h}}^{\mathrm{max}}$ scenario)~\cite{Benchmarks} and one in
which no mixing in the scalar top sector is assumed.  The excluded
domains in the [\mh,\tanb] plane are shown in Fig.~\ref{fig:nn_tanb}
as obtained using the NN stream.  The results in the
$m_{\mathrm{h}}^{\mathrm{max}}$ scenario are summarized in
Table~\ref{nn_summarytable}.  The \tanb\ exclusion range is sensitive
to the top quark mass assumption which was taken to be 175\,\Gcs. For
a top quark mass of 180\,\Gcs, the \tanb\ exclusion range reduces to
[1.0,1.5].

For the invisibly decaying Higgs boson search, no discriminant 
variable is used in the confidence level calculations, as was done in
the previous search at $188.6\,\G$~\cite{189inv}, because the sliding
analysis technique effectively acts as if the reconstructed mass were
used as a discriminant variable.  The results are interpreted as an
exclusion domain in the (\mh,$\xi^2$) plane, presented in
Fig.~\ref{plan}. For $\xi^2 = 1$, invisibly decaying Higgs bosons with
masses below $106.4\,\Gcs$ are excluded at $95\%$ CL with an expected
limit of $105.6\,\Gcs$.

\begin{table}[t]
\caption{\capstyl Summary of the $95\%$ CL lower limits on Higgs 
boson masses in the Standard Model and in the MSSM, and 
their expected values in absence of signal.  The $\tan\beta$
excluded ranges in the $m_{\mathrm{h}}^{\mathrm{max}}$ scenario
are also given for a top quark mass of $175\,\Gcs$.
\label{nn_summarytable}}
\begin{center}
\begin{tabular}{|c||c|c||c|c|}
\hline\hline
  & \multicolumn{4}{|c|}{$95\%$ CL lower limits (\Gcs)} \\ 
\cline{2-5}
   & \multicolumn{2}{|c||}{NN stream} & \multicolumn{2}{|c|}{Cut stream} \\
\cline{2-5} 
       & Median Expected & Observed & Median Expected & Observed \\
\hline
SM \mh & 107.8 & 107.7 & 107.3 & 105.2 \\
\hline
MSSM \mh & 88.7 & 91.2 & 88.3 & 90.7 \\
MSSM \mA & 89.1 & 91.6 & 88.7 & 91.1 \\
\hline
$\tan\beta$ & [0.8,1.9] & [0.8,1.9] & [0.8,1.9] & [0.8, 1.7]\\
\hline\hline
\end{tabular}
\end{center}
\end{table}

\vspace{-0.5cm}
\section{Conclusions}
\label{summary} 

Searches for neutral Higgs bosons in \epem\ collisions at
centre-of-mass energies from 191.6 to 201.6\,\G\ have been carried out
with the \ALEPH\ detector using an integrated luminosity of
237\,\invpb.  The major event topologies expected from the \h\Z\
process for visibly as well as invisibly decaying Higgs bosons, the
weak boson fusion process, and the \h\A\ process have been analysed.
The data were combined with the 176\,\invpb\ collected in 1998 at a
centre-of-mass energy of 188.6\,\G.

A $95\%$ CL lower limit on the Standard Model Higgs boson mass of
107.7\,\Gcs\ is obtained with an expected sensitivity of 107.8\,\Gcs.
Lower limits of 91.2 and 91.6\,\Gcs\ are obtained at $95\%$ CL for
the masses of the h and A neutral Higgs bosons of the MSSM.  These
limits, determined in the $m_{\mathrm{h}}^{\mathrm{max}}$ scenario,
are valid for any $\tan\beta>0.5$.  For a top quark mass of 175\,\Gcs,
the \tanb\ range between 0.8 and 1.9 is excluded with $95\%$
confidence.

An invisibly decaying Higgs boson with a production cross section
equal to that in the Standard Model is excluded at
$95\%$ CL for masses below 106.4\,\Gcs.

\newpage 
\section*{Acknowledgements}
We wish to congratulate our colleagues from the accelerator divisions
for the very successful operation of \LEP\ at high energies.  We are
indebted to the engineers and technicians in all our institutions for
their contribution to the excellent performance of \ALEPH.  Those of
us from non-member countries thank \CERN\ for its hospitality.

\end{document}